\documentclass[prl,twocolumn,amsmath,amssymb,superscriptaddress,preprintnumbers,floatfix,letterpaper]{revtex4}

\usepackage{graphicx,psfrag,amsmath,color}
\usepackage{times}

\newcommand{\bk}{{\bf k}}

\newcommand{\br}{{\bf r}}

\newcommand{\half}{\frac{1}{2}}

\begin{document}

\title{Emergence of Cooper pairs, d-wave duality
and the phase diagram of cuprate superconductors}
\author{Zlatko Te\v sanovi\'c}
\affiliation{Department of Physics and Astronomy, Johns Hopkins
University, Baltimore, MD 21218, USA 
\\ {\rm(\today)}
}

\maketitle

{\bf 
BCS theory describes the formation of 
Cooper pairs and their instant ``Bose condensation'' into 
a superconducting state. Helium atoms are preformed bosons and, 
in addition to their condensed superfluid state, can also form a 
quantum solid, lacking phase-coherence. Here we show that 
the fate of Cooper pairs can be more varied than the BCS 
or helium paradigms. In copper-oxide d-wave 
superconductors (dSC) Cooper pairs are non-local 
objects, with both center-of-mass and relative motions. 
As doping decreases, the center-of mass fluctuations force a correlated
dSC into a state with enhanced diamagnetism and 
robust but short-ranged superconducting order. At extreme underdoping, 
the relative fluctuations take over and two pseudogaps -- 
``small'' (charge) and ``large'' (spin) -- emerge naturally from the theory, 
as Cooper pairs ``disintegrate'' and charge ``detaches'' 
from spin-singlet bonds.  The ensuing 
ground state(s) are governed by antiferromagnetic rather than 
by superconducting correlations. The theory is used to account
for recent experiments and to draw 
general conclusions about the phase diagram.
}

Recent experiments  \cite{fischer} 
have narrowed the field of contenders
for theoretical description of high-$T_c$ cuprates
while simultaneously promoting the nature of the pseudogap state to
the key conceptual issue.
Two basic ideas are in play \cite{fischer}: either the pseudogap is a quantum
disordered d-wave superconductor, 
or an entirely different form of a ``competing'' order, 
originating from the particle-hole (diagonal) channel 
\cite{kivelson,nagaosa,zaanen,ddw}.

In this paper, I show that the pseudogap in a correlated 
lattice dSC is {\em both}. Several recent experiments
offer important clues in this respect:
the observation of enhanced
quantum diamagnetism in the pseudogap state of LSCO \cite{luli},
alongside the giant Nernst effect \cite{luli}, 
points to an intimate relation between the pseudogap and a
{\em quantum disordered} d-wave superconductor (dSC) \cite{melikyan} -- 
there is hardly another known
microscopic mechanism which can deliver 
diamagnetism of this magnitude. Furthermore, 
the angle-resolved photoemission (ARPES) in underdoped LBCO \cite{valla}
reveals a nodal d-wave type excitation spectrum
inside the pseudogap, testifying to the shared origins with the 
superconducting state. Importantly however, the observed
quantum diamagnetism \cite{luli} terminates at very low but
{\em finite} underdoping and thereupon the
magnetic signal is dominated by spin response. This is
in line with the ARPES data on BSCCO \cite{zx},
infrared ellipsometry on RBCO \cite{keimer} and
the most recent STM data \cite{hudson}, which are suggestive of
two pseudogaps: the nodal one, 
reminiscent of a superconductor and decreasing at extreme underdoping,
and the antinodal one, which appears to be dominated by spin correlations
and remains large.

Two kinds of new results are reported: First, 
guided by broadly accepted
microscopic features of cuprates, a
theory for the long-distance charge $2e$ sector of 
the pseudogap state is constructed and shown to
provide quantitative understanding
of the observations in \cite{luli}, 
including the measured upper critical field at $T=0$
and the {\em quantum} vortex liquid and solid at lower
fields. Furthermore, 
a dSC dome in the doping-magnetic field-temperature ($x$-$H$-$T$) phase
diagram is found to be enveloped by a 
larger, {\em charge $2e$} Cooper pairing dome, 
dominated by {\em quantum} fluctuations of the superconducting
order parameter $\Psi$. 
This larger dome  {\em collapses} to $T=0$ at {\em finite} $x=x_0\sim 0.01$, 
while the amplitude of the microscopic 
{\em spin-singlet} pairing term $\Delta_{jk}$ remains large as $x\to 0$.
Between $x_0$ and $x_c\sim 0.055$, the ground state is the
quantum disordered dSC, the order in  $\Psi$ 
($\sim \langle e^{i\theta_{jk}}\rangle$) preempted
by free vortex-antivortex excitations of $|\Delta_{jk}|e^{i\theta_{jk}}$.
Two distinct pairing energy scales -- 
$\Psi\Delta$ 
for the charge and $\Delta$ for the spin sector --
provide a natural
explanation for ``small'' and ``large'' pseudogaps observed
by various experimental probes \cite{fischer,zx,keimer,hudson}.

Second, a sharp distinction is drawn between 
the dSC (off-diagonal or particle-particle)
and particle-hole (diagonal) regimes of the microscopic theory
in terms of the {\em symmetry of the action} for 
the {\em bond} phase $\theta_{jk}$ of $\Delta_{jk}$: 
in the former case this symmetry is of the 
XY type, while in the latter it is related 
to a compact gauge symmetry, and thus much larger. 
A quantum phase transition between the
two regimes is likely at extreme underdoping $x\lesssim x_0$.
The XY regime is dominated by the center-of-mass motion
of the Cooper pairs, while the relative
motion fuses into low energy physics in the ``compact gauge''
regime, with fundamental consequences.
By analogy with its more familiar s-wave 
(bosonic) cousin \cite{fisherlee,melikyan,franzduality,zaanen},
we dub the above sequence of events -- unleashed by intensifying
quantum fluctuations of $\theta_{jk}$ on approach to the
Mott insulating state -- a ``d-wave duality''.

The starting point is a general strongly correlated
microscopic Hamiltonian which almost certainly underlies the essential
physics of cuprate superconductors:
\begin{multline}
\hat H= -\sum_{\langle ij\rangle,\sigma}c^\dag_{i\sigma}\hat t^*c_{j\sigma}
+ \sum_{\langle ij\rangle}\hat\Delta_{ij}[c^\dag_{i\uparrow}c^\dag_{j\downarrow}
- c^\dag_{i\downarrow}c^\dag_{j\uparrow}] 
+ ({\rm h.c.})
\\ 
+\frac{2}{{\cal J}_{\rm eff}}\sum_{\langle ij\rangle}\hat\Delta_{ij}^\dag\hat\Delta_{ij}
+ \sum_i U_p\hat n_{i\uparrow}\hat n_{i\downarrow} + \sum_{(i,j)}W_{ij}\hat n_i\hat n_j~.
\label{hamiltonian}
\end{multline}
In (\ref{hamiltonian}) $c^\dag_{i\sigma}$ ($c_{i\sigma}$) 
creates (annihilates) electrons of spin $\sigma$ on site $i$
of the CuO$_2$ lattice,
$\hat n_{i\sigma}\equiv c^\dag_{i\sigma}c_{i\sigma}$,
$\hat n_i\equiv \hat n_{i\uparrow}+ \hat n_{i\downarrow}$,
$\hat t^*$ is an effective hopping, 
$\hat\Delta_{ij}$ is the Hubbard-Stratonovich
``pairing'' operator, the integration over 
which yields the superexchange term
${\cal J}_{\rm eff}\sum_{\langle ij\rangle}[\hat{\bf S}_i\cdot\hat{\bf S}_j
-\frac{1}{4}\hat n_i\hat n_j]$, and
$U_p$ is an arbitrarily large repulsion, a purely mathematical tool
introduced to suppress double occupancy
\cite{hubbard}. $\{W_{ij}\}$ are the
extended-range (nearest, next-nearest,...)
Coulomb and possibly phonon-induced interactions. At half-filling 
($x=0$), $\hat n_i\equiv 1$ and
all but the ``pairing'' terms in (\ref{hamiltonian}) 
drop out; the result is the $S=\half$ Heisenberg Hamiltonian 
and the ground state is the Neel-Mott antiferromagnet (AF).
At $x>0$, the precise nature of the ground state of strongly interacting
Hamiltonians (\ref{hamiltonian}) is unknown  but we assume
a domain of its parameters where a correlated BCS-type d-wave 
superconductivity is stabilized, as established 
in a myriad of experiments \cite{fischer}.

We anchor our analysis
to the off-diagonal
saddle point of (\ref{hamiltonian}):
$|\Delta_{jk}|=\Delta_d=\pm\Delta$ at $x(y)$ bonds.
As $x$ decreases toward $x_c$ and beyond, the increasing quantum disorder
in $\theta_{jk}$ progressively destabilizes 
the dSC ground state.
Nevertheless, we persist in following this progression all the way to the 
Heisenberg AF at $x=0$, which, in a sense precisely defined below,
 is an ``{\em infinitely}'' strongly
quantum fluctuating dSC.

To follow this path of growing fluctuations in 
 $\theta_{jk}$, we integrate out the fermions in
(\ref{hamiltonian}) about the above saddle point.
This is (much!) easier said than done: the presence of gapless nodal
fermions in a d-wave superconductor
makes the resulting action severely non-analytic.
Still, as long as we stay focused on physics intrinsically
tied to Cooper pairs and steer clear of
nodal quasiparticles (or gapless spinons discussed in \cite{nagaosa}),
the following ({\em bond}) phase-only action for the correlated
dSC captures the essentials:
${\cal S}_{XY}^d=\int {\cal D}\varphi_i\exp\bigl( - \int d\tau {\cal L}_{XY}^d\bigr)$, where
\begin{multline}
{\cal L}_{XY}^d =
i\sum_if_i\dot\varphi_i+ \frac{\kappa _0}{2}\sum_i\dot\varphi_i^2 
-J\sum_{nn}\cos(\varphi_i - \varphi_j) 
\\
-J_1\sum_{rnnn}\cos(\varphi_i - \varphi_j)
-J_2\sum_{bnnn}\cos(\varphi_i - \varphi_j)
\\
-\sum_{\hskip .02 truein \square}K\cos(\square\theta)
+{\cal L}_{\rm nodal}[\cos(\varphi_i - \varphi_j)]+{\cal L}_{\rm core}~.
\label{lagrangian}
\end{multline}
In (\ref{lagrangian}), $\{\varphi_i\}$ are
{\em identical} to the {\em bond} phases
$\{\theta_{jk}\}$ of $\{\hat\Delta_{jk}\}$, with subscript $i$
referring to
the center of each bond $\langle jk\rangle$, the first term is the charge $2e$
Berry phase, $nn$ and $r(b)nn$ are nearest and 
(inequivalent!) next-nearest neighbors on the $\{i\}$ lattice,
$\square$ indicates plaquettes of the CuO$_2$ lattice, 
and $\square\theta\equiv\theta_{12}-\theta_{23}+\theta_{34}-\theta_{41}$
around each plaquette. The meaning of (\ref{lagrangian})
and its relation to (\ref{hamiltonian}) are
detailed in \cite{melikyan} and in Fig. 1. We will use
(\ref{lagrangian}), and lean on (\ref{hamiltonian}) when necessary,
to analyze the low-energy physics of cuprates.

\begin{figure}[tbh]
\centering
\includegraphics[viewport=90 520 520 750,scale=0.1,width=0.4\columnwidth]{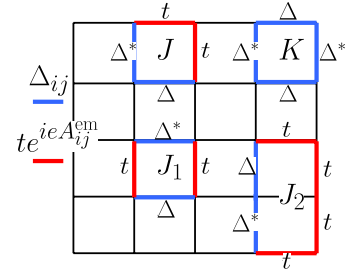}

\vskip .6truein
\caption{\label{phaseaction} A sampler of processes in
(\ref{hamiltonian}) contributing to various couplings
in (\ref{lagrangian}). The vector potential ${\bf A}^{\rm em}$
couples only to hopping $t^*$.
}
\end{figure}
To extract the long distance effective theory of
{\em quantum} fluctuations in 
$\{e^{i\varphi_i}\equiv e^{i\theta_{jk}}\}$, we introduce two
{\em real} Hubbard-Stratonovich fields, 
$\{\Pi_i(\tau)\}$ and $\{{\rm Q}_i(\tau)\}$, to decouple terms like
$\exp\bigl(J\sum_{\langle i,j\rangle}\cos(\varphi_i-\varphi_j)\bigr)$
in (\ref{lagrangian}) as
\begin{multline}
\prod_i\int d\Pi_i d{\rm Q}_i
\exp\bigl[2\sum_i \bigl(\Pi_i\cos\varphi_i+{\rm Q}_i\sin\varphi_i\bigr)-
\\
\sum_{i,j}\bigl(\Pi_i,{\rm Q}_i\bigr)
\langle i|(J\sum_{\alpha=x,y}\cos k_\alpha a)^{-1}|j\rangle
\bigl(\Pi_j,{\rm Q}_j\bigr)^{\rm T}\bigr]~.
\label{HS}
\end{multline}
Upon introduction of the complex field 
$\Psi_i\equiv \Pi_i +i{\rm Q}_i$
coupled to $e^{-i\varphi_i}$,
the integration over $\{\varphi_i(\tau)\}$ yields
\begin{multline}
\int {\cal D}\Psi_i(\tau)\exp\bigl[
\int \frac{d\tau}{\tau_v}\sum_i{\cal B}(|\Psi_i(\tau)|^2)-
\\
\int\frac{d\tau}{\tau_v}\int\frac{d\tau '}{\tau_v}\sum_{i,j}
\Psi^*_i(\tau)
\langle i,\tau|{\cal G}_0(\bk,\omega)|j,\tau\prime\rangle
\Psi_j(\tau ')
\bigr]
\label{psiaction}
\end{multline}
\noindent
as the representation of (\ref{lagrangian}) in terms of $\{\Psi_i(\tau)\}$.
In (\ref{psiaction}),
${\cal B}(|z|^2)\equiv
\ln\bigl[\int_{-\pi}^{\pi}d\varphi \exp (z e^{-i\varphi }+ z ^*e^{i\varphi})\bigr]$
and
${\cal G}_0^{-1}(\bk,\omega)$ is a $2\times 2$ matrix,
with diagonal elements 
$J_{1}\cos(k_xa)+J_{2}\cos(k_ya)+\frac{1}{2}\kappa_0\omega^2$ and
$J_{2}\cos(k_xa)+J_{1}\cos(k_ya)+\frac{1}{2}\kappa_0\omega^2$,
and off-diagonal ones $J[\cos((k_x-k_y)a/2)+\cos((k_x+k_y)a/2)]$.
The $2\times 2$ form of ${\cal G}_0^{-1}(\bk,\omega)$ simply reflects 
$J_1\not =J_2$ and the resulting inequivalence of the nearest
neighbor sites of the $\{i\}$ lattice, centered on bonds of the
original CuO$_2$ lattice (Fig. 1).

Several simplifications are built into (\ref{psiaction}) in
anticipation of our next step: 
$K$ (\ref{lagrangian})
is assumed to be small in the
regime of interest \cite{melikyan} (see later text, however). 
Furthermore, at long distances relevant to \cite{luli}, we can
neglect the Berry phase in (\ref{lagrangian}) which induces
$2e$ superstructure modulations in CuO$_2$ lattice
at intermediate lengthscales. $\tau_v$, the vortex tunneling
time, is a short time cutoff: $\tau_v\sim m_va^2$,
where  $m_v$ is the (anti)vortex mass,
equal to only a few electron masses in underdoped cuprates
\cite{melikyan,sachdevmass}. $\tau_v$ should not change dramatically within
the window of $x$ probed in \cite{luli}.

Assuming $|\Psi_i(\tau)|^2\ll 1$,
focusing on 
long distances and low energies, and
restoring ${\bf H}$
via ${\bf\nabla}\to{\bf\nabla}-i(2e){\bf A}$
gives
${\cal S}_{XY}^d\to{\cal S}_{QGL}^d=\int d(\tau/\tau_v) d^2(\br/a) {\cal L}_{QGL}^d[\Psi(\br,\tau)]$, where
\begin{equation}
{\cal L}_{QGL}^d = \gamma_\tau |\dot\Psi|^2 +
\gamma |(\nabla -i(2e){\bf A})\Psi|^2 + 
\alpha |\Psi|^2 + \frac{1}{4}|\Psi|^4~,
\label{QGL}
\end{equation}
$\gamma_\tau = \kappa _0/8J_d^2\tau_v^2$, 
$\gamma = a^2\tilde J/8J_d^2\tau_v^2$, $\alpha = (1/2J_d\tau_v) -1$, 
$J_d=J+\half J_1+\half J_2$, 
and
$\tilde J=J+J_1+J_2$. Note that, consistent with our strategy,
${\cal L}_{\rm nodal}$ (\ref{lagrangian}) produces only higher (third order,
non-analytic) derivatives in (\ref{QGL}).

The physical meaning of (\ref{QGL}) is important:
${\cal L}_{QGL}^d$
represents the low energy, long-distance charge $2e$ sector
of (\ref{hamiltonian}) in the regime of its parameters
and doping characterized by strong {\em quantum} superconducting
fluctuations. While the procedure that led us to (\ref{QGL})
from (\ref{lagrangian}), via (\ref{HS}, \ref{psiaction}), 
bears a superficial resemblance to the standard Gor'kov's derivation
of the Ginzburg-Landau functional from the BCS theory, the actual
physics is {\em different}: by setting $\Psi$ to a complex constant
in the standard Gor'kov-Ginzburg-Landau functional and focusing on the
quadratic term, one is probing the {\em weak-coupling}
instability of the electronic system
toward {\em formation} of Cooper pairs, i.e. a finite complex $\Delta$,
accompanied by full phase coherence.
Since the pairing susceptibility of a typical
normal system diverges as $T\to 0$,
such instability is always present as long as there is an attractive 
coupling constant in a given angular momentum channel. 
In contrast, $\Psi$ in (\ref{QGL}) is testing 
for {\em phase order} in $e^{i\theta_{jk}}$
in the system in which large $|\Delta_{jk}|\not = 0$ is already established.
The susceptibility to such phase order is typically {\em finite}, something
the reader can easily check by using (\ref{HS}) and
applying our procedure to the simple
quantum XY model \cite{overhead}. 
This fact of much theoretical importance constitutes 
the foundation of various dual descriptions of 
{\em strongly correlated} systems \cite{fisherlee,melikyan,franzduality}:
a single effective action (\ref{lagrangian}) describes both ordered and
quantum disordered ground states -- depending on the strength of its
coupling constants -- just like the original microscopic Hamiltonian.
Obviously, (\ref{QGL}) as it follows from (\ref{lagrangian})
is only a coarse-grained caricature
of (\ref{hamiltonian}). Nevertheless, its key features faithfully
reflect general properties of the underlying microscopic theory.

The first among these are Mott correlations 
which, as $x\to 0$, suppress double
occupancy and drive kinetic energy toward zero.
This implies $J_d(x)\to 0$: it becomes
impossible to establish phase coherence between $\Delta_{ij}$
on different bonds (Fig. 1). This is reflected in
$\alpha(x\to 0)\to +\infty$ (\ref{QGL}).
In the language of (\ref{QGL}) $x=0$ acts as an infinite ``temperature''
and thus the familiar Heisenberg AF at half-filling 
is ``infinitely far'' removed from
a dSC (and vice versa!). 
This statement is made more precise 
from the standpoint of the microscopic theory (\ref{hamiltonian}) 
later in the text.

While the exact calculation of $t^*(x)$, $\Delta_d(x)$,
and $J_d(x)$ is beyond reach, the above trend is manifest
in various approximations ($t^*\sim xt$, $\Delta_d\sim {\cal J}_{\rm eff}$
as $x\to 0$ \cite{kotliar,plainvanillarvb,nagaosa}). 
In any case, the precise form of $J_d(x)$
is {\em not needed} for our considerations: we can expand
$\alpha (x)\approx \alpha_0 (x_0 - x)$ about
$\alpha (x=x_0)= (1/2J_d(x_0)\tau_v(x_0)) -1=0$, the 
above arguments clearly implying that $x_0$ is {\em finite}.
A reasonable estimate gives $x_0 \sim \Delta_d /t$, where
$t$ is the {\em bare} hopping.

With these simplifications, we use (\ref{QGL}) to address
certain basic features of the cuprates' $x$-$H$-$T$ phase diagram. In 
uniform $H$, we follow Abrikosov and expand
the quadratic terms in (\ref{QGL}) in 
charge $2e$ Landau levels (LLs) (distinct from
single electron LLs!) 
$\Psi(\br,\tau)=\sum_j \Psi_j(\br,\tau)$, $j=0,1,2,\dots$:
\begin{equation}
{\cal L}_{GL}^d \to
\sum_{j=0}^{\infty}\bigl[\gamma_\tau|\dot\Psi_j |^2 +
\alpha_j(x,H)|\Psi_j|^2\bigr] + \frac{1}{4}|\Psi |^4~,
\label{lls}
\end{equation}
where $\alpha_j= \alpha_0\bigl(x_0-x + (2j+1)H/H^\prime_{c2}\bigr)$.
$\alpha_{j=0}(x,H) = 0$ determines the upper critical field
\begin{equation}
H_{c2}(x,T\to 0)=H^\prime_{c2}(x-x_0)~,~~H^\prime_{c2}=\alpha_0/4\pi e\gamma~.
\label{hc2}
\end{equation}
In Fig. 2(a) we compare (\ref{hc2}) to \cite{luli} and find
good agreement with the choice $x_0\approx 0.01$ and 
$H^\prime_{c2}\approx 9.4~{\rm Tesla}/\%$ of doping.

The upper critical field (\ref{hc2})
is a mean-field concept -- it decribes the line
in the $H$-$x$ phase diagram along which
${\cal S}_{QGL}^d [\Psi (\br,\tau)]$ 
(\ref{QGL}) has an absolute minimum 
for a {\em single} configuration of 
$\Psi (\br,\tau)= \Psi_{\rm mf}(\br)$. 
At $H=0$,
$\Psi_{\rm mf}=\sqrt{2\alpha_0(x-x_0)}\not =0$,
implying $\langle e^{i\theta_{ij}}\rangle\not = 0$,
for $x>x_0$. For $x<x_0$, however, the minimum of 
${\cal S}_{QGL}^d$ is 
at $\Psi (\br,\tau)=\langle e^{i\theta_{ij}}\rangle =0$! 
Thus, for low $x$, the above
mean-field theory of ${\cal S}_{XY}^d$ (\ref{lagrangian}) and, by
inference, of its microscopic parent Hamiltonian (\ref{hamiltonian}),
is intrinsically {\em different} from {\em both} the 
standard BCS theory
{\em and} the slave boson mean-field approximation
and the related
Gutzwiller-projected BCS wavefunction 
\cite{kotliar,plainvanillarvb,nagaosa}.

\begin{figure}[tbh]
\centering
\includegraphics[viewport=70 520 520 700,scale=0.3,width=0.75\columnwidth]{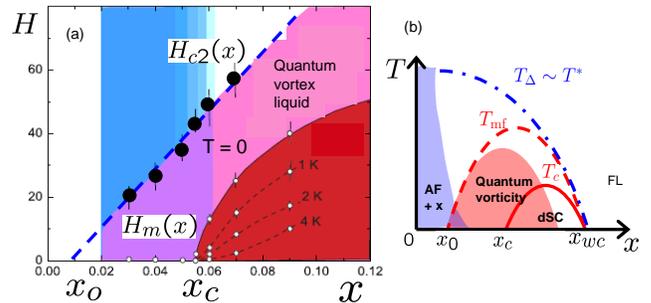}
\vskip .5truein
\caption{\label{luliphasediagram} (a) $H$ [Tesla] vs. $x$
phase diagram of LSCO adapted from \cite{luli}. All data
points are from \cite{luli}. Thick (blue) dashed line is 
$H_{c2}(x,T\to 0)$ (\ref{hc2}), with $x_0=0.01$ and
$H_{c2}^\prime \sim 9.4~{\rm Tesla}/\%$. $H_m(x,T\to 0)$ is the
boundary of the dark (red) region. $H_m(x)$ turns convex
by $T\sim 4K$, in line with (\ref{hc2lll},\ref{hc2xy}). 
The lighter (blue) region shows increasing spin response \cite{luli}; (b)
 $x$-$T$ phase diagram of cuprates
based on (\ref{lagrangian},\ref{QGL}). Note that $T_{\rm mf}$ is roughly
of the order of the ``small'' charge $2e$ pseudogap $\Psi_{\rm mf}\Delta_d$
while $T_\Delta\sim T^*$ corresponds to the ``large'' pseudogap $\Delta_d$.
}
\end{figure}

This difference stems from an important physical information captured
by (\ref{QGL}): In a BCS-style weak-coupling theory, it is the kinetic
energy part of (\ref{hamiltonian}) that is large at the
point of the quantum phase transition, while the amplitude
$\Delta_d\to 0$. Thus, the moment the self-consistent
solution $\Delta_d\not =0$ appears, the phase coherence between
different bonds is automatically
established. $\Psi_{\rm mf}$ is immediately 
finite and proportional to
$\Delta_d$. Incidentally, this is the likely 
physics behind the closing of the dSC
dome on the {\em overdoped} side of the phase diagram in Fig. 2(b), at
$x=x_{wc}$. Beyond $x_{wc}$, there are
{\em no} superconducting fluctuations in the ground state; the system
is in its underlying {\em normal} state, most likely the Fermi liquid.



In the {\em underdoped} regime, near $x_0$, the situation is 
reversed. At low $x$, we are dealing with rapidly growing 
Mott correlations and the ensuing suppression of coherent motion
of $2e$ charge. Now, it is the kinetic energy that is
becoming small while the {\em spin-singlet} pairing 
$|\Delta |$ remains large. 
However, 
the superconductivity
now does {\em not} appear the moment $t^*\not =0$ -- in 
a strongly correlated
system the kinetic energy must 
rise to a {\em finite} fraction of
$\Delta_d$ {\em before} even a mean-field solution $\Psi_{\rm mf}\not =0$
becomes possible at $x=x_0$. This crucial feature of the
microscopic theory (\ref{hamiltonian}) is faithfully captured
by Eq. (\ref{QGL}) and the mean-field approximation
$\Psi(\br,\tau)\to\Psi_{\rm mf}$. For $x <x_0$, even though 
$\Delta_d$ remains large ($\sim {\cal J}_{\rm eff}$), 
$\Psi_{\rm mf}$ 
collapses to zero (Fig. 2), reflecting enormous quantum fluctuations in
$\{\theta_{ij}\}$,
unrestrained by the coherence between different bonds.
In contrast, the slave boson and
the Gutzwiller-projected mean-field 
theories \cite{kotliar,plainvanillarvb,nagaosa}
produce a dSC the moment kinetic energy, $\sim xt$,
becomes finite. 

How are the above mean-field arguments generalized to $T\not = 0$? 
Various coefficients in (\ref{QGL})
have temperature dependences that are difficult to determine directly
from a strongly interacting theory (\ref{hamiltonian}).  
Again, a simple Gaussian correction to the quadratic
term in (\ref{lls})
$\alpha_j \to \alpha_j(x,H,T)
\approx \alpha_0\bigl(x_0-x + (2j+1)H/H^\prime_{c2} + T^2/\Theta^2\bigr)$
will suffice for our needs and gives
\begin{equation}
H_{c2}(x,T)=H^\prime_{c2}(x-x_0 - T^2/\Theta^2)~.
\label{hc2t}
\end{equation}
A corollary of (\ref{hc2t}) is that, at $H=0$, 
$T_{\rm mf}=\Theta\sqrt{x-x_0}\ll T_{\Delta}\sim\vert\Delta\vert$ defines the line in the $x$-$T$ phase
diagram below which 
$\Psi_{\rm mf} (x,T)=\sqrt{2\alpha_0(x-x_0-T^2/\Theta^2)}$ becomes
finite (Fig. 2(b)). 
The estimate from the 
underdoped side of ``Nernst dome'' \cite{luli}
yields $\Theta \sim 37 K/\sqrt{\%}$ in LSCO.

The above mean-field transition to
$\Psi_{\rm mf}(\br)\not =0$ from $\Psi_{\rm mf}(\br)=0$
is only a crossover: Eqs. (\ref{hc2},\ref{hc2t})
define the underdoped side of the ``Nernst dome'' (or
``diamagnetic dome'') in the $x$-$H$-$T$ phase diagram \cite{luli} along
which enhanced  diamagnetism and (anti)vortex fluctuations become
observable (see Fig. 2). These are but a couple of manifestations of the rise
of superconducting correlations and of $|\Psi(\br,\tau)|$ in (\ref{QGL}) 
sampling large local values even in absence of the long-range
off-diagonal order.
Actually, much is known about the fluctuations in 
(\ref{QGL}, \ref{lls}) (see \cite{GLzbt,sudbo} and references therein). 
At $H=0$, the quantum phase transition to the true superconducting ground state
is shifted to $x_c$ from $x_0$: $x_c-x_0\sim {\rm Gi}$, where ${\rm Gi}$
is the dimensionless coefficient of the quartic term in (\ref{QGL})
\cite{GLzbt}; ${\rm Gi}\sim 0.045$ from our fit to
\cite{luli}. At finite $T$, the true superconducting transition line
emanates from $x_c$ as
\begin{equation}
T_c(x)\sim \rho_s\sim \lambda^{-2}_s\sim 
\chi_d^{z/(z-2)}\sim \xi^{-z}_{\rm sc}\sim |x-x_c|^{z\nu}~ , 
\label{tc}
\end{equation}
where $\rho_s (x)$, $\lambda_s(x)$, $\chi_d(x)$,
$\xi_{\rm sc}(x)$, $\nu$, and $z$ are
the superfluid density, penetration depth, diamagnetic susceptibility,
correlation length, its exponent, and dynamical
exponent, respectively, all associated with
the {\em quantum} superconducting transition at $x_c$. 
Eq. (\ref{QGL})
predicts $\nu=\nu_{\rm XY}\sim 0.667$ and $z=1$ but a 
more detailed treatment \cite{melikyan},
incorporating the Berry phase (\ref{lagrangian})
can lead to different values. The regime between the mean-field crossover and
the true transition is dominated by {\em quantum}
vortex-antivortex fluctuations (Fig. 2(b)) \cite{footnotekt}.

Similarly, the true transition for $H\not =0$ in (\ref{QGL}) takes
place only far below $H_{c2}(x,T)$ (\ref{hc2t}), at $H_m(x,T)$. 
The $x$-$H$-$T$ phase diagram between $H_{c2}(x,T)$ 
and $H_m(x,T)$ is occupied by the {\em quantum} vortex liquid (Fig. 2(a)), 
exhibiting strong vortex fluctuations and
enhanced diamagnetism. Along $H=H_m(x,T)$, this quantum
vortex liquid freezes into a solid.
and any weak pinning ensures superconductivity. 
The shape of the $H=H_m(x,T)$ line depends
on whether one is in the ``low'' or ``high'' field regime of
(\ref{lls}) \cite{GLzbt}. For $H\gg H_b$,
the quartic term in (\ref{lls}) is ineffective in mixing the LLs
of the quadratic part; in the opposite limit, $H\ll H_b$, this mixing
is strong and the $H\not = 0$ behavior is dictated by the quantum
vortex-antivortex unbinding at $H=0$ \cite{GLzbt,sudbo}. An
estimate \cite{GLzbt} gives 
$H_b\sim ({\rm Gi}/13) H^\prime_{c2}\sim 4~{\rm Tesla}$ in LSCO,
about midway through the fluctuation region explored in \cite{luli}.

At high fields, $H\gg H_b$, (\ref{lls}) is dominated by
the lowest Landau level (LLL) ($j=0$).
${\cal S}_{QGL}^d\to {\cal S}^d_{QGL-LLL}=\int d(\tau/\tau_v) d^2(\br /a){\cal L}_{QGL-LLL}^d[\Psi_{0}(\br,\tau)]$, where: 
\begin{equation}
{\cal L}_{GL-LLL}^d =
\gamma_\tau|\dot\Psi_{0} |^2 +
\alpha_{j=0}(x,H,T)|\Psi_{0}|^2 + \frac{1}{4}|\Psi_{0}|^4~,
\label{QGL-LLL}
\end{equation}
with $\alpha_{j=0}= \alpha_0\bigl(x_0-x + H/H^\prime_{c2}+T^2/\Theta^2\bigr)$.
The functional integral $\int {\cal D}\Psi\exp(-{\cal S}^d_{GL-LLL})$ is
confined to the LLL, 
$\Psi \in \Psi_{0}(\br,\tau)=\Phi(\tau)\prod_i(z-z_i(\tau))\exp(-|z|^2/4)$,
where $z=(x+iy)/\ell$,  $\ell = \sqrt{1/2eH}$ and
$\{z_i(\tau)\}$ are the quantum zeroes (vortices) of 
$\Psi(\br,\tau)$ \cite{GLzbt}. A prominent
feature of (\ref{QGL-LLL}) is that the physical quantities computed
from it exhibit {\em single-parameter} scaling: ground state energy,
magnetization, low $T$ thermodynamics, etc., are all {\em universal}
functions of 
$g(x,H,T)={\rm const.}\times(x_0-x+ H/H^\prime_{c2}+T^2/\Theta^2)/H^{2/3}$ only. Various scaling functions are computed in \cite{GLzbt}. 
In particular,
the quantum vortex liquid-solid transition in $\{z_i(\tau)\}$
observed in \cite{luli} takes place at 
$g(x,H,T)=g_m<0$, where $g_m$ is a universal number, $g_m\approx -3$.
This leads to 
\begin{equation}
H_m(x,T)\sim (x-x_0-T^2/\Theta^2)^{3/2}~.
\label{hc2lll}
\end{equation}
$H_{c2}(x,T)$ (\ref{hc2}) corresponds to $g=0$;
thus, the fluctuation region in this regime
expands to $H_{c2}-H_m \sim {\rm Gi}^{2/3}$.

In the opposite limit, $H\ll H_b$, the $H=0$ quantum transition governs scaling
properties: the universal scaling functions 
depend on $\xi_{\rm sc}(x,T=0)/\ell$
and  $H_m(x)$ emerges from $x_c$ as 
\begin{equation}
H_m(x)\sim (x-x_c)^{2\nu}~.
\label{hc2xy}
\end{equation}

Note that both (\ref{hc2lll}) and (\ref{hc2xy})
are convex lines while the observed $H_m(x,T\to 0)$ is 
concave (Fig 2(a)). The likely cause is pinning of quantum vortices
by the CuO$_2$ lattice (or disorder), absent from
continuum theory (\ref{QGL}). This explanation is supported by the 
$H_m(x)$ curves in \cite{luli} turning convex at a finite, but still low, $T$
($\sim 4~K\ll|\Delta |$) suggestive of thermal depinning
and the $H_m(x)$ lines for different $T$
{\em all} terminating at the same $x_c\approx 0.055$ as $H\to 0$.
The number of data points in \cite{luli} is insufficient for a detailed
fit to (\ref{hc2lll},\ref{hc2xy}) -- it would be interesting to test these
predictions in further experiments.

The physics captured by (\ref{QGL}) and discussed above
reflects the center-of-mass motion of
Cooper pairs. To display this more clearly consider slowly changing
phase $\theta_{jk}$, far from the cores of vortex or antivortex defects. 
Since electrons live on the {\em sites} of 
CuO$_2$ lattice it is convenient to define two sets of site variables:
phases $\{\phi_j\}$
as $e^{i\phi_j}\equiv\sum_k e^{i\theta_{jk}}/|\sum_k e^{i\theta_{jk}}|$
where $k$ runs over four bonds emanating from site $j$, and
complex numbers $\{s_\omega\}$ as
$s_\omega\equiv e^{i\theta_{12}}-e^{i\theta_{23}}+e^{i\theta_{34}}-e^{i\theta_{41}}$ where $\omega$ denotes a ``virtual'' site at the center of
each $\langle 1234\rangle$ plaquette of the CuO$_2$ lattice. Up to and
including second order derivatives one can invert these definitions
to find 
$e^{i\theta_{jk}}=e^{i\theta^{\rm CM}_{jk}+i\theta^{\rm r}_{jk}}$,
where 
$e^{i\theta^{\rm CM}_{jk}}=(e^{i\phi_j}+e^{i\phi_k})/|e^{i\phi_j}+e^{i\phi_k}|$ \cite{index}  and $\theta^{\rm r}_{jk}[\{\phi_j\},\{s_\omega\}]$ is the
``irreducible'' part of $\theta_{jk}$ which cannot be reduced to site
phases. By inserting the above expression for $\theta_{jk}$ into
(\ref{hamiltonian}) and simply repeating the steps that 
led from (\ref{lagrangian}) to (\ref{QGL}), one arrives 
at precisely the same phase structure as in (\ref{QGL}), 
except that now $\dot\phi(\br,\tau)$
and $\nabla\phi(\br,\tau) -(2e){\bf A}$ replace the corresponding
derivatives of $\Psi$. Evidently,
$\phi_j$ plays the role of the overall center-of-mass 
phase of a Cooper pair. Importantly, $\theta^{\rm r}_{jk}$
does not appear at the second order level in derivatives and is
naturally interpreted as corresponding to
the relative motion of the paired
spin-singlet bonds in (\ref{hamiltonian}).

This brings us to the second feature that transpires from (\ref{QGL}) 
and concerns extreme underdoping $x<x_0$:
in this regime $\alpha\to +\infty$ and off-diagonal correlations
are maximally suppressed. Such limit of extreme quantum
fluctuations is {\em different} for the d-wave {\em bond} phase 
$\theta_{jk}$ than for
the site phase in an s-wave (bosonic) XY duality
\cite{fisherlee,franzduality}. In the s-wave case, the dual opposite
of a superconductor is a Wigner crystal of Copper pairs.
This remains an entirely center-of-mass affair, with the 
quantum disorder in the corresponding site phase promoted
by suppressed density fluctuations of positionally ordered
Cooper pairs. But off-diagonal correlations are still strong, as
exemplified by quantum diamagnetism and vortex-antivortex
fluctuations, and the ground state is still well described by
the XY-like phase action; these properties are broadly
similar to the intermediate
``Quantum vorticity'' state in Fig. 2(b) although there are important
differences as well, like the presence of spinful nodal fermions
in the latter.

By contrast, in the d-wave case, the extreme quantum 
disorder limit ($x\to 0$) of the 
action (\ref{lagrangian}) does
{\em not} have the XY symmetry: $t^*\to 0$ and the XY parts of
(\ref{lagrangian}) disappear, leaving behind only the plaquette-type 
action, since terms like
$K$ include only $\Delta$'s (Fig. 1). This demise of the XY behavior
in (\ref{lagrangian}) and (\ref{hamiltonian}) -- 
occasioned by Mott suppression of kinetic energy --
translates into collapse of off-diagonal correlations and is identified here
as the microscopic physics behind the observed ``two pseudogaps'' in
underdoped cuprates \cite{zx,keimer,hudson}, as alluded to
earlier in the text (Fig. 2(b)).  To see this explicitly,
we consider below the off-diagonal, ``anomalous'' 
electron propagator following
from (\ref{hamiltonian}):
$F(\br-\br',\tau -\tau')= \langle c^\dag_\uparrow(\br,\tau)c^\dag_\downarrow (\br ',\tau ')\rangle$. Finite $F$ implies superconducting off-diagonal
long-range order.

We can actually compute $F$ with the accuracy needed to make 
contact with the recent experiments \cite{zx,keimer,hudson}. First,
within the ``Quantum vorticity'' state in Fig. 2(b) it suffices to keep
only the leading relevant derivatives and we can replace 
$e^{i\theta_{jk}}\to e^{i\theta^{\rm CM}_{jk}}=(e^{i\phi_j}+e^{i\phi_k})/|e^{i\phi_j}+e^{i\phi_k}|$,
as explained above. Next, we define new fermions $f_{i\sigma}$
as $c_{i\sigma}=f_{i\sigma}e^{\frac{i}{2}\phi_i}$, 
$c^\dag_{i\sigma}=e^{-\frac{i}{2}\phi_i}f^\dag_{i\sigma}$. This 
leads to:
\begin{equation}
F(\br-\br',\tau -\tau')\approx \langle\Psi\rangle\langle f^\dag_\uparrow(\br,\tau)f^\dag_\downarrow (\br ',\tau ')\rangle~~,
\label{anomalous}
\end{equation}
accurate up to leading derivatives, where $\langle\Psi\rangle$ is nothing but
the expectation value of the superconducting order parameter
computed from (\ref{QGL}). The precise structure of the
off-diagonal $f$-fermion propagator (\ref{anomalous}) is a complex
problem, as $f$ fermions behave as neutral d-wave quasiparticles
{\em interacting} via gauge fields generated by
spacetime derivatives of the phase 
$\half\phi$ \cite{melikyan}; nevertheless, for
the purposes of overall energetics and at the
level of resolution available in \cite{zx,keimer,hudson}, such
propagator behaves simply as -- a somewhat smeared version of -- the ordinary
BCS d-wave anomalous propagator. Furthermore, Refs. \cite{zx,keimer,hudson}
are sensitive only to short range off-diagonal order
and thus $\langle\Psi\rangle$
should be replaced by $\Psi_{\rm mf}$. After the Fourier transform we
finally obtain:
\begin{equation}
F(\bk,\omega)\approx \Psi_{\rm mf}
\frac{\Delta_{\hat\bk}}{-\omega^2 + ({\bf v}_F\cdot (\bk -\bk_F))^2 + |\Delta_{\hat\bk}|^2}~~,
\label{anomalousfinal}
\end{equation}
where $\Delta_{\hat\bk}=\Delta\cos (2\phi_\bk)$, $\phi_\bk$ is
the angle going around the Fermi surface and ${\bf v}_F$ and
${\bf k}_F$ are the Fermi velocity and momentum, respectively.
From (\ref{anomalousfinal}) it follows naturally that 
$\Psi_{\rm mf}\Delta$ should be interpreted as the ``small'' pseudogap,
measuring charge $2e$ off-diagonal correlations and collapsing near $x_0$,
while the spin-singlet pairing gap $\Delta$ itself remains large (Fig. 2(b)).

The above collapse of ``small'' pseudogap
$\Psi_{\rm mf}\Delta$ at $x<x_0$ implies
that vortex-antivortex pairs, the
staple of XY-type physics, are displaced as relevant excitations.
To see what replaces them, we transform
$c_{i,\sigma}\to \sigma d^\dag_{i,-\sigma}$ on one of the sublattices
of the CuO$_2$ lattice,
turning the spin-singlet pairing
$|\Delta | e^{i\theta_{jk}}c^\dag_{j,\sigma}c^\dag_{k,-\sigma}$
(\ref{hamiltonian}) into hopping 
$|\Delta |e^{i\theta_{jk}}c^\dag_{j,\sigma}d_{k,\sigma}$.
After this transformation,
$\{\pm\theta_{jk}\}$ on alternating bonds are
cast in the role of a gauge field
$\{a_{jk}\}$ coupled to the {\em staggered} ``charge'', switching
from $+1$ to $-1$ between the two sublattices. 
In this language,
the plaquette terms in (\ref{lagrangian}) are nothing but
the action for $\{a_{jk}\}$, 
invariant under the compact gauge transformations
$a_{jk}\to a_{jk}+\zeta_j-\zeta_k$.
This is a much larger symmetry than the XY one of 
(\ref{lagrangian}) at finite $x$ and $t^*$. 
The relevant excitations of a compact gauge theory
are monopoles and antimonopoles in $\{a_{jk}\}$ \cite{nagaosa} -- in turn,
monopoles in $\{a_{jk}\}$
are closely related to the {\em relative}
motion of Cooper pairs and the bond phases $\{\theta^{\rm r}_{jk}\}$
defined earlier. This is easily appreciated by observing
that the plaquette term in (\ref{lagrangian}) always equals
$\pm 1$ when $\theta_{jk}$ is restricted to $\theta^{\rm CM}_{jk}$ alone.
Only by allowing unrestricted fluctuations of $\theta_{jk}$ and
including the relative phases $\theta^{\rm r}_{jk}$, can one successfully
proliferate monopoles and antimonopoles in $a_{jk}$.
With (\ref{hamiltonian})
serving as the microscopic model, the monopoles will be
in their plasma phase and thus
cofinining. Actually, despite apparent mathematical opaqueness of
our argument, this is just one among many 
disguises of the traditional Heisenberg-Mott AF \cite{nagaosa}.
This opaqueness is an unavoidable consequence of our persistence in following
the fate of a dSC all the way to the half-filling, the route along
which the quantum
fluctuations in the phase $\theta_{jk}$ of its gap function progressively
become enormous and ultimately
unrestrained. In this sense, it is the standard Heisenberg-Mott
AF that stands at the far end ($x=0$)
of the ``d-wave dual road'' for correlated dSC \cite{herbut}.

How is the XY symmetry recovered at finite $x$ and $t^*$?
The $J_1$ term in (\ref{lagrangian}) provides an
illustration (Fig. 1):
it moves charge $2e$ ``Cooper pair'' from one 
spin-singlet bond to the next, a process
vital for establishing local superconducting phase coherence. 
In the gauge theory language this is a Higgs term: 
$\cos(\theta_{ij}-\theta_{kl})\to\cos(a_{ij}+a_{kl})$!
Thus, the charge $2e$ motion and the associated kinetic energy 
correspond to the symmetry breaking Higgs terms in the gauge theory
dialect. The presence of Higgs terms
suppresses free monopoles in $\{a_{ij}\}$ and promotes its gauge
sector to physical reality -- the result is the 
{\em conservation of vorticity} in $\{\theta_{ij}\}$ and the XY symmetry
of its action (\ref{lagrangian}).
Thus, Copper pairs ``emerge'' from the underlying
dynamics of (\ref{hamiltonian}) within the ``Quantum vorticity'' region
in Fig. 2(b), as {\em dual partner} of well-defined
vortex-antivortex excitations.
One is tempted to speculate that
a quantum transition between the ``compact'' 
and ``XY'' regimes takes place at {\em finite}
$x\sim x_0$, with the AF+$x$ ground state(s) in Fig. 2(b) regulated by
monopoles and antimonopoles in $a_{jk}$ and reflecting the relative motion,
instead of vortices and antivortices associated with the Cooper pair
center-of-mass fluctuations in $\theta_{jk}$. Such ground states
are the true ``competing orders'' of a dSC, 
arising from the particle-hole (diagonal)
channel, and likely include non-uniform AF and stripes \cite{kivelson,zaanen}.


I thank N. P. Ong, L. Li, M. Franz, A. Melikyan, S. Sachdev, A. Sudb\o , and
O. Vafek for useful discussions.
This work was supported in part by the NSF grant DMR-0531159.



\begin{references}
\bibitem{hubbard} In particular, $U_p$ is arbitrarily larger than the
Hubbard $U$, already used to generate superexchange
${\cal J}_{\rm eff}\sim t^2/U$ in (\ref{hamiltonian}). Related
to this, the precise form
of $\hat t^*$ depends on the method one uses to suppress doubly
occupied sites. For example, if doubly occupied sites are simply
projected out from the Hilbert space of (\ref{hamiltonian}) one can
set $\hat t^*\to\hat t^*_\sigma\to (1-\hat n_{i,-\sigma})t(1-\hat n_{j,-\sigma})$. If
one uses the ``slave boson'' method \cite{nagaosa}, then
$\hat t^*\to \hat r_it\hat r_j$, where $\hat r_i$ is the slave boson creation (annihilation)
operator stripped of its phase and $\hat n_i + \hat r_i^2=1$.
Irrespective of its precise form,
the only property that matters for the purposes of this paper
is that $\hat t^*$ involves only electron densities $\hat n_{i\sigma}$.
\bibitem{overhead} Actually, one can apply an equivalent procedure to 
(\ref{hamiltonian}) directly, but at the cost
in mathematical overhead that is too large for this paper.
\bibitem{footnotekt} The finite $T$ transition itself is of the Kosterlitz-Thouless kind and reverts
to the XY universality class once the interlayer coupling between 
CuO$_2$ planes is included.
\end{references}
\end{document}